\documentclass{statsoc}
\usepackage[utf8]{inputenc}
\usepackage[english]{babel}
\usepackage{amsmath}
\usepackage{amssymb}
\usepackage{amsfonts}
\usepackage{graphicx}
\usepackage{natbib}
\usepackage[margin=2cm]{geometry}

\title[Multiple seeds and disconnected networks in RDS]{Multiple seed structure and disconnected networks\\in respondent-driven sampling}
\author[J.\ Malmros, L.\ Rocha]{Jens Malmros\thanks{\textit{Address for correspondence:} Jens Malmros, Department of mathematics, Stockholm University, SE-106 91, Stockholm, Sweden,\\\textsf{E-mail: jensm@math.su.se}}}
\address{Stockholm University,
Sweden}
\author[J.\ Malmros, L.\ Rocha]{Luis E.C.\ Rocha}
\address{Karolinska Institutet,
Stockholm,
Sweden\\
Université de Namur,
Belgium}

\DeclareMathOperator{\prob}{\mathbb{P}}
\DeclareMathOperator{\expectation}{\mathbb{E}}
\newcommand{\overbar}[1]{\mkern 1.5mu\overline{\mkern-1.5mu#1\mkern-1.5mu}\mkern 1.5mu}

\begin{document}

\maketitle

\begin{abstract}
Respondent-driven sampling (RDS) is a link-tracing sampling method that is especially suitable for sampling hidden populations. RDS combines an efficient snowball-type sampling scheme with inferential procedures that yield unbiased population estimates under some assumptions about the sampling procedure and population structure. Several seed individuals are typically used to initiate RDS recruitment. However, standard RDS estimation theory assume that all sampled individuals originate from only one seed. We present an estimator, based on a random walk with teleportation, which accounts for the multiple seed structure of RDS. The new estimator can also be used on populations with disconnected social networks. We numerically evaluate our estimator by simulations on artificial and real networks. Our estimator outperforms previous estimators, especially when the proportion of seeds in the sample is large. We recommend our new estimator to be used in RDS studies, in particular when the number of seeds is large or the social network of the population is disconnected.
\end{abstract}

\section{Introduction}\label{Sec:Introduction}

Some human populations are difficult to survey for various reasons, for example, if no sampling frame for the population exists and the population is small relative to the general population, if members of the population are difficult to identify or unwilling to disclose themselves, or if individuals in the population are reluctant to participate in surveys. Examples of such \emph{hidden} or \emph{hard-to-survey} populations \citep{schwartlaender2001,tourangeau2014} include several groups that are subject to marginalisation or stigmatisation, e.g., injecting drug users, homosexual men, sex workers, illegal immigrants, and the homeless \citep{beyrer2012,faugier1997,sudman1991}. Because of their characteristics, hidden populations can often not be satisfactorily investigated using standard sampling procedures and thus alternative sampling and estimation techniques must be considered \citep{Magnani2005,barros2015}. A reasonably efficient and cost-effective way to sample from hidden populations is to utilise link-tracing techniques \citep{thompson1990,thompson2000,ThompsonSampling}. In such procedures, the population is assumed to be connected by a social network and previously sampled individuals are engaged in the recruitment of their social contacts to the sample. While link-tracing procedures have been considered relatively efficient in collecting sufficiently sized samples from hidden populations, the samples obtained have often been viewed as convenience samples not suitable for inference because of the substantial bias that occurs from the selection procedure \citep{erickson1979}.

A relatively recent and increasingly popular link-tracing methodology is \emph{respondent-driven sampling} (RDS) \citep{heckathorn1997}. The method is essentially an extension of snowball sampling \citep{biernacki1981} for which inferential procedures facilitating unbiased population estimates have been developed \citep{salganik2004,volz2008}. An RDS study begins with the formation of an initial group of individuals, the \emph{seeds}, which are typically recruited among known population members. The seeds are provided with coupons, typically between three to five, which are to be distributed to their peers in the population of interest. An individual that has received a coupon is eligible for participation in the study upon presenting the coupon at the study site. After taking part in the study, sampled population members (i.e., respondents) are also given coupons which are to be distributed to those of their peers which have not yet participated in the study. This is repeated with subsequently sampled individuals until the desired sample size has been reached or until recruitment ceases by itself, in which case often additional seeds are recruited among not yet sampled members of the target population in order to re-initiate recruitment to the sample \citep{malekinejad2008}. There are typically incentives given to individuals both for their own participation as well as for the participation of those to whom they have given coupons.

The most commonly used RDS estimator, the Volz-Heckathorn (V-H) estimator \citep{volz2008}, assumes that the RDS recruitment process can be approximated by a simple random walk on the social network of the population and also makes some assumptions about the structure of the social network. For example, it is assumed that sampling occurs with replacement, that respondents recruit randomly from their social contacts, and that the social relations in the population are mutual. It is also assumed that respondents accurately report their number of social contacts, or, in network lingo, their \emph{degree}. In the general with-replacement design-based sampling framework, we can form an asymptotically unbiased Hansen-Hurwitz type ratio estimator \citep{hansen1943} of the mean of a population trait $y$ from a sample $S$ as
\begin{equation}\label{Eq:RDSEstimator}
 \hat\mu = \frac{\sum_{u\in S}\frac{y_u}{p_u}}{\sum_{u\in S}\frac{1}{p_u}},
\end{equation}
where $y_u$ and $p_u$ are the values of $y$ and the draw-wise selection probability for a sampled individual $u$, respectively. For the V-H estimator, $\hat\mu_{\rm{V-H}}$, $p_u$ is replaced by $d_u$, since population members are assumed to be sampled with probability proportional to their degree from the random walk in stationarity in this case. V-H estimates have been shown to be sensitive to situations where the relatively strong assumptions on the recruitment process and the network structure do not hold \citep{Gile2010SocMet,Goel2010,Lu2012JRSS,McCreesh2012,Tomas2011EJS,Wejnert2009}. In part due to these results, other RDS estimators have been developed \citep{Gile2011JASA,LuEtal,Lu2013,Gile2015ModelAssisted}.

\begin{figure}
\centering
\includegraphics[scale=0.8]{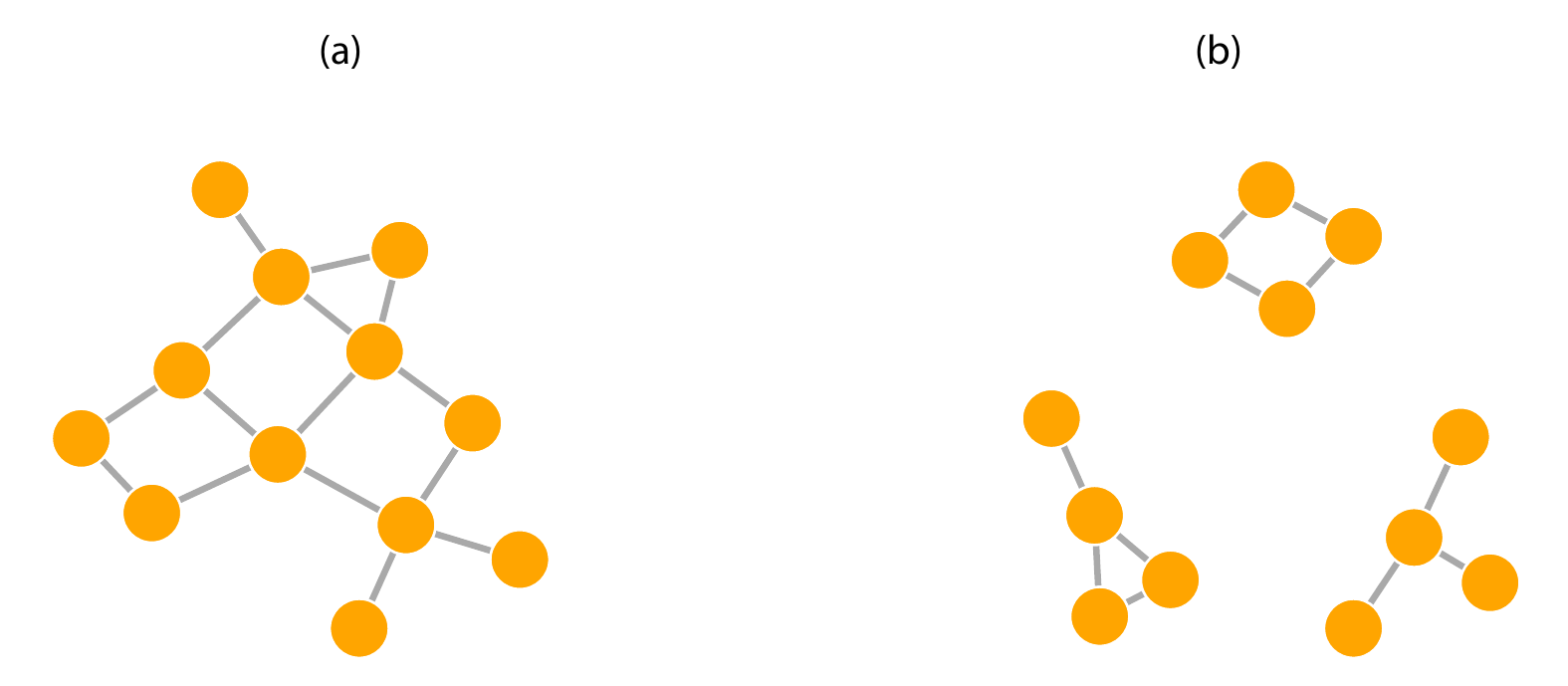}
\caption{Schematic illustration of two social networks where the circles represent the vertices and the lines represent the edges. In (a), the network is connected in one component. In (b), the network is disconnected with three connected components.}\label{Fig:NetworkIllustration}
\end{figure}

When the social network of the population is \emph{connected}, the social relationships between individuals bind together all members of the population in one connected \emph{component} (Figure~\ref{Fig:NetworkIllustration}~a). Otherwise, the network is \emph{disconnected} and has several connected components (Figure~\ref{Fig:NetworkIllustration}~b). It is in the nature of the simple random walk that it can not reach all population members in populations with disconnected social networks. This means that the V-H estimator is not applicable to such populations; a shortcoming that is shared with all proposed RDS estimators available in the literature. Additionally, the structural properties of the network may be such that a link-tracing sampling procedure is contained within parts of the network, something which is likely to affect RDS estimates even though the network is connected \citep{burt2010,mccreesh2011,salganik2006}. This may e.g.\ be the result of community structure within the network, where groups of individuals are more closely connected among each other than with individuals in different groups \citep{rocha2016} or so-called bottlenecks, where a single individual is the link between otherwise disconnected parts of the network \citep{johnston2013}. Most RDS studies start with several seeds (e.g., 10) that each initiates a recruitment tree of its own. However, the simple random walk approximation assumes that only one recruitment chain can be used to describe the whole sample. This discrepancy becomes larger as the proportion of seeds in the sample increases, which may, e.g., be the result of additional seeds being recruited to the study. This is not uncommon; for example, in \citep{malekinejad2008}, 43\% of the reviewed empirical RDS studies with available data reported the use of additional seeds and in several studies, the sample consisted of more than 10\% seeds. In some cases, the proportion of seeds in the sample may be as large as 18\% \citep{stein2014} or even 30\% \citep{stromdahl2015}.

In this paper, we extend RDS estimation to account for the multiple seed structure and populations with disconnected social networks. We use a \emph{random walk with teleportation} (RWWT) to model the RDS recruitment process \citep{brin1998}. The RWWT may, in each time step, go to a randomly chosen social contact of the currently visited individual, like the simple random walk used in the V-H estimator, or jump to a randomly chosen individual in the population. Hence, the RWWT may visit parts of the social network that are not connected to each other and explore them separately, which is not possible in the simple random walk. Moreover, the set of individuals visited by the RWWT will be made up of multiple chains of neighbouring population members, each originating from a randomly selected individual (among all individuals). Each of these chains may be viewed as an approximation to the recruitment tree originating from one seed. Hence, the RWWT is able to account for the multiple seed structure of RDS under the assumption that seeds are selected uniformly; this assumption is discussed in Section~\ref{Sec:Discussion}. To describe the social structure of our target population, we use the so-called \emph{configuration model}~\citep{Molloy1995,Molloy1998}, a random graph model for which the degree distribution of the resulting graph may be specified, to fit the network of interest.

The rest of the paper is organised as follows. In Subsections~\ref{Subsec:NetworkModel} and \ref{Subsec:RWWT} we formally define the configuration model and the RWWT, respectively. In Section~\ref{Sec:Calculations}, we present our calculations for the stationary distribution of the process (Subsection~\ref{Subsec:StatDist}) and how to estimate it (Subsection~\ref{Subsec:EstimationOfParameters}). We evaluate our estimator by simulations for varying proportion of seeds in the sample and for populations with disconnected networks in Section~\ref{Sec:Simulations}. Our findings are then discussed in Section~\ref{Sec:Discussion}.

\section{Preliminaries}

We first introduce some network terminology which is used in the following. Formally, a social network is composed of a set of \emph{vertices} $V$ that represents the actors (e.g., individuals) and a set of edges $E$ which represents the relations that connect the actors together (see Figure~\ref{Fig:NetworkIllustration}). The network can be represented by its \emph{adjacency matrix} $A=\{a_{uv}\}$, where $u$ and $v$ belong to the set of vertices. We consider \emph{undirected} networks only, i.e., networks where all relations are mutual. Then, $a_{uv}=a_{vu}=1$ if there is an edge between two vertices $u$ and $v$ and $a_{uv}=a_{vu}=0$ otherwise. We say that two vertices $u$ and $v$ are \emph{neighbours} if there is an edge between $u$ and $v$. As previously mentioned, the degree $d_u$ of a vertex $u$ is the number of contacts of $u$, where $d_u=\sum_va_{uv}=\sum_va_{vu}$.

\subsection{Configuration model}\label{Subsec:NetworkModel}

Assume that we have a set of $n$ vertices. For the results in the rest of the paper, we consider the infinite population limit $n\to\infty$. Let $D$ be a random variable, defined on the non-negative integers, that represents the degree distribution, i.e., the distribution of vertex degrees. To construct the network, we assign to each vertex a number of stubs or half-edges, independently drawn from $D$. Then, we randomly form pairs of all the stubs. If the number of stubs is uneven, we discard one stub, which does not affect our results in the limit of infinite population size. This construction procedure may generate self-loops and multiedges, i.e., edges that connects a vertex to itself and several edges between the same two vertices; the proportion of these is however small when $\expectation(D^2)<\infty$. In particular, the probability that the generated graph is simple, i.e., that it contains no self-loops or multiedges, is bounded away from 0 if $\expectation(D^2)<\infty$ \citep[e.g.,][Lemma 5.3]{britton2007}. Hence, we condition on the generated graphs being simple under the assumption that the second moment of $D$ is finite in the following. We denote networks generated from this model by $G(V,E)$, where $V$ is the set of vertices and $E$ is the set of edges.

\subsection{Random walk with teleportation}\label{Subsec:RWWT}

A RWWT in discrete time $\{Z_t;t=0,1,2,\ldots\}$, taking place on a network $G(V,E)$, is a Markov process with state space given by the vertex set of the network. In each step, the walker traverses to a randomly chosen neighbour of the last visited vertex with probability $c\in[0,1]$ or moves to a uniformly chosen vertex $v\in V$ with probability $1-c$. Let the transition probability between two vertices $u$ and $v$ be $p_{uv}$. The transition matrix $P=\{p_{uv}\}$, $u,v\in V$, of $\{Z_t\}$ is then given by
\[P=cAD^{-1} + (1-c)\frac{1}{n}\bar{1}^T\bar{1},\]
where $A$ is the adjacency matrix of $G$, $D$ is a diagonal matrix with the degree sequence of vertices in $G$ at its diagonal, and $\bar{1}$ is the column vector of ones.

\section{Theory}\label{Sec:Calculations}

\subsection{Stationary distribution}\label{Subsec:StatDist}

Assume that we have a configuration model network $G(V,E)$ of size $n$, where $n$ is assumed to be large. Let the degree distribution of $G$ be given by the random variable $D$. Further assume that we have a RWWT $\{X_t;t=0,1,2,\ldots\}$ taking place on this network. We assume that the structure of $G$ is unknown in $\{X_t\}$ but that the degrees of visited vertices are known. Let $v\in V$ be an arbitrarily chosen fixed vertex with known degree $d_v$. We are interested in the limiting probability that $X_t$ is at $v$.

Assume that the random walk visits vertex $u\ne v$ at time $s$. In what follows, we write $u\leftrightarrow v$ if $u$ and $v$ are neighbours and $u\nleftrightarrow v$ otherwise. The probability that $v$ is visited at time $s+1$ is then given by
\begin{align*}
p_{uv} &= \prob(X_{s+1}=v|X_s=u)\\
&= \prob(X_{s+1}=v|u\leftrightarrow v,X_s=u)\prob(u\leftrightarrow v)\\
&+ \prob(X_{s+1}=v|u\nleftrightarrow v,X_s=u)\prob(u\nleftrightarrow v).
\end{align*}
First, we consider the case where $u$ and $v$ are neighbours. Let $J$ denote the event that the random walk makes a random jump at $s$. We have
\begin{align*}
\prob(X_{s+1}=v|u\leftrightarrow v,X_s=u) &= \prob(X_{s+1}=v|u\leftrightarrow v,X_s=u,J)\prob(J)\\
&+ \prob(X_{s+1}=v|u\leftrightarrow v,X_s=u,J^\complement)\prob(J^\complement).
\end{align*}
By the definition of $\{X_t\}$, $\prob(X_{s+1}=v|u\leftrightarrow v,X_s=u,J)=1/n$ and $\prob(J)=1-c$. If the random walk does not jump at $s$, it will only visit $v$ at $s+1$ if it traverses along the edge between $u$ and $v$, which happens with probability $1/d_u$. Hence,
\[\prob(X_{s+1}=v|u\leftrightarrow v,X_s=u) = \frac{1}{n}(1-c) + \frac{1}{d_u}c.\]
If $u$ and $v$ are not neighbours, $v$ may only be visited at $s+1$ through a random jump. Hence,
\begin{align*}
\prob(X_{s+1}=v|u\nleftrightarrow v,X_s=u) &= \prob(X_{s+1}=v|u\nleftrightarrow v,X_s=u,J)\prob(J)\\
&= \frac{1}{n}(1-c).
\end{align*}
By construction, $\prob(u\leftrightarrow v)=d_ud_v/(2|E|-1)$, where we may approximate $2|E|-1$ by $2|E|=n\expectation(D)$ for large $n$. From these results, we have for $u\ne v$ that
\begin{align*}
p_{uv} &\approx \left(\frac{1}{n}(1-c) + \frac{1}{d_u}c\right)\frac{d_ud_v}{n\expectation(D)} + \frac{1}{n}(1-c)\left(1-\frac{d_ud_v}{n\expectation(D)}\right)\\
&= c\frac{d_v}{n\expectation(D)} + (1-c)\frac{1}{n}\\
&= \frac{1}{n}\left(c\frac{d_v}{\expectation(D)}+1-c\right).
\end{align*}
If we the random walk visits $v$ at time $s$, then it may only visit $v$ again at $s+1$ by a random jump; hence,
\[p_{vv}=\frac{1}{n}(1-c).\]
Define
\[\pi_u=\frac{1}{n}\left(c\frac{d_u}{\expectation(D)}+1-c\right)\]
for all $u\in V$. We have
\begin{align*}
\sum_{u\in V}\pi_up_{uv} &= \sum_{u\in V;u\ne v}\pi_up_{uv}+\pi_vp_{vv}\\
&= \frac{1}{n}\left(c\frac{d_v}{\expectation(D)}+1-c\right)(1-\pi_v)+\frac{1}{n}(1-c)\pi_v\\
&= \frac{1}{n}\left(c\frac{d_v}{\expectation(D)}+1-c\right)-\frac{cd_v}{n\expectation(D)}\pi_v\\
&\approx \pi_v,
\end{align*}
where the approximation comes from that $cd_v/(n\expectation(D))\pi_v$ is $O(1/n^2)$. Because $v$ was arbitrarily chosen,
\begin{equation}\label{Eq:pi_v}
\pi_v=\frac{1}{n}\left(c\frac{d_v}{\expectation(D)}+1-c\right)\propto c\frac{d_v}{\expectation(D)}+1-c, \hspace{10pt}v\in V
\end{equation}
gives the stationary distribution of the RWWT on the configuration model network.

Note that, because the transition probabilities and the stationary distribution are very similar, it is close to redundant to make the assumption that the process is in stationarity when we later consider sampling from this process; this was also noted in \citet{Gile2011JASA} for the simple random walk on the configuration model network. We will however do so for a stringent exposition. Also note that it has been shown that the RWWT has the same stationary distribution on Chung-Lu random graphs and Erd\H{o}s-Rényi random graphs \citep{Kadavankandy2015}. In general however, there exists no closed expression for the stationary distribution of the RWWT on an undirected network \citep{Grolmusz2015}. A generalisation of the RWWT is to let the probability to be visited when a jump has occurred to be different between different vertices, which in general will introduce a dependence on $n$ in Eq.~\eqref{Eq:pi_v}. If however the probability that a vertex is visited when a jump has occurred is proportional to its degree, all individuals are sampled with probability proportional to degree and the V-H estimator is recovered.

\subsection{Estimation of $c$ and $\expectation(D)$}\label{Subsec:EstimationOfParameters}

Under the assumption that we obtain our sample by sampling with replacement from the RWWT in stationarity, we can use the stationary distribution from Eq.~\eqref{Eq:pi_v} as the draw-wise selection probabilities in the estimator in Eq.~\eqref{Eq:RDSEstimator}. However, in order to do so, we need to estimate the unknown parameter $c$ and $\expectation(D)$. From here on we assume that $S$ is a sample of size $n_S$ from an RDS study with $m$ seeds in which, for each sampled individual $u$, the property of interest $y_u$ and the degree $d_u$ is recorded. Under the assumptions of Subsection~\ref{Subsec:StatDist}, we may view this sample as the outcome of a RWWT on a configuration model network which has jumped at $m$ occasions during the collection of the sample. The jump probability $1-c$ can then be estimated by the proportion of seeds in the sample $m/n_S$ and we get an estimator $\hat c$ of $c$ as
\begin{equation}\label{Eq:chat}
\hat c=1-\frac{m}{n_S}.
\end{equation}
In order to estimate $\expectation(D)$, we consider a partition of the sample $S$ in two parts: $S_J$ which consists of those individuals that were sampled as the result of a jump by the random walk and $S_{RW}$ which consists of those individuals that were sampled as the result of an edge traversal. Because the inclusion of an individual in either partition of $S$ are independent of the composition of the other partition under our assumptions, $S_J$ and $S_{RW}$ are independent. The sample partitions are easily identified from the RDS sample; $S_J$ comprises the seeds and $S_{RW}=S\setminus S_j$.  The sizes of $S_J$ and $S_{RW}$ are given by $m$ and $n_S-m$, respectively. We will proceed by deriving two estimators $\widehat{\expectation(D)}_J$ and $\widehat{\expectation(D)}_{RW}$ of the expected degree from sampled individuals in $S_J$ and $S_{RW}$, respectively. The individuals in $S_J$ are sampled randomly with replacement and hence an estimator of $\expectation(D)$ is \citep[][ch.\ 2.9]{sarndal1992}
\begin{equation}\label{Eq:EDJ}
\widehat{\expectation(D)}_J=\frac{\sum_{u\in S_J}d_u}{m}.
\end{equation}
The variance of $\widehat{\expectation(D)}_J$ is estimated by
\begin{equation}\label{Eq:VarJ}
\widehat{Var}\left(\widehat{\expectation(D)}_J\right)=\frac{s_J^2}{m},
\end{equation}
where $s_J^2$ is the sample variance of the degrees of individuals in $S_J$. Because the individuals in $S_{RW}$ are sampled by edge traversal in the random walk, their draw-wise selection probabilities are proportional to their degree. We have that an asymptotically unbiased estimator of the expected degree can be derived from the ratio of two Hansen-Hurwitz estimators \citep{salganik2004} as
\begin{equation}\label{Eq:EDRW}
\widehat{\expectation(D)}_{RW}=\frac{n_S-m}{\sum_{u\in S_{RW}}1/d_u}.
\end{equation}
We obtain an approximative estimator of $Var\left(\widehat{\expectation(D)}_{RW}\right)$ by applying the Delta method and substituting population quantities with their sample counterparts:
\begin{equation}\label{Eq:VarRW}
\widehat{Var}\left(\widehat{\expectation(D)}_{RW}\right)\approx \left(\frac{1}{\left(\overbar{d^{-1}}\right)^4}\right)\frac{s_{d^{-1}}^2}{n_S-m},
\end{equation}
where $\overbar{d^{-1}}$ and $s_{d^{-1}}^2$ are the sample mean and variance of the inverse degrees of individuals in $S_{RW}$, respectively. Then, we combine these estimators in a composite estimator $\widehat{\expectation(D)}$ \citep{schaible1978} of the expected degree:
\begin{equation}\label{Eq:CompositeEstimator}
\widehat{\expectation(D)}=w\widehat{\expectation(D)}_J + (1-w)\widehat{\expectation(D)}_{RW},
\end{equation}
where $0\le w\le1$. We want to choose $w$ such that the variance of $\widehat{\expectation(D)}$ is minimized. Because $S_J$ and $S_{RW}$ are independent samples, the variance of $\widehat{\expectation(D)}$ is a weighted sum of the variances of $\widehat{\expectation(D)}_J$ and $\widehat{\expectation(D)}_{RW}$. Taking the variance and differentiating in Eq.~\eqref{Eq:CompositeEstimator} yields that the minimal variance is obtained when $w=w^*$, where
\begin{equation}\label{Eq:OptimalWeight}
w^*=\frac{Var\left(\widehat{\expectation(D)}_{RW}\right)}
{Var\left(\widehat{\expectation(D)}_J\right)+Var\left(\widehat{\expectation(D)}_{RW}\right)}.
\end{equation}
We obtain an estimate $\hat w^*$ by substituting the estimates from Eqs.~\eqref{Eq:VarJ} and \eqref{Eq:VarRW} into Eq.~\eqref{Eq:OptimalWeight}. To find estimates $\{\hat\pi_u;u\in V\}$ of the stationary distribution of the RWWT on the configuration model network we may then substitute the estimates given by Eqs.~\eqref{Eq:chat}, \eqref{Eq:CompositeEstimator}, and \eqref{Eq:OptimalWeight} into Eq.~\eqref{Eq:pi_v}; we have
\begin{equation}\label{Eq:pi_uEstimate}
\hat\pi_u\propto\hat c\frac{d_u}{\widehat{\expectation(D)}}+1-\hat c, u\in S.
\end{equation}
The estimated stationary distribution can then be substituted into Eq.~\eqref{Eq:RDSEstimator} to obtain an estimator $\hat\mu_{\text T}$ of population properties as
\begin{equation}\label{Eq:EstRDSEstimator}
\hat\mu_{\text T}=\frac{\sum_{u\in S}\frac{y_u}{\hat\pi_u}}{\sum_{u\in S}\frac{1}{\hat\pi_u}}.
\end{equation}
From Eqs.~\eqref{Eq:pi_v}, \eqref{Eq:pi_uEstimate}, and \eqref{Eq:EstRDSEstimator}, we recover, as limiting cases for $\hat\mu_{\text T}$, the sample mean when $c\to0$, i.e., when the draw-wise selection probabilities all are similar, and the V-H estimator when $c\to1$.

\section{Numerical simulations}\label{Sec:Simulations}

Our estimator extends the V-H estimator in two respects: i) it accounts for the multiple seed structure of RDS and ii) it is valid for disconnected networks. We focus on these properties of the estimator in our evaluation and compare with the limiting cases given by the V-H estimator and the sample mean. We do not consider estimators that require population parameters that are traditionally not collected or estimable within the RDS sample \citep{Gile2011JASA,LuEtal,Lu2013,Gile2015ModelAssisted}.

To test the performance of our estimator, we simulate the RDS process in a population represented by a random network. We generate the network with $N=10000$ individuals (i.e.\ the size of the target population) using the configuration model in which the degree distribution $D$ is given by $\prob(D=d)=p_d$. We consider two degree distributions: a power-law with exponential cutoff, for which
\[p_d =  \dfrac{\lambda^{1-\alpha}}{\Gamma(1-\alpha,\lambda d_{\text{min}})} d^{-\alpha} \exp(-\lambda d),\]
and a log-normal, for which
\[p_d =  \dfrac{1}{d \sigma\sqrt{2\pi}} \exp\left(-\dfrac{(\ln \; d - \theta)^2}{2 \sigma^2}\right).\]
These distributions are chosen because they reproduce the degree heterogeneity observed in social networks \citep[e.g.,][]{Amaral2000}. We choose the parameters $d_{\text{min}}=3$, $\alpha=2.5$, and $\lambda=0.00001$ for the power-law, and $\theta=2.0$ and $\sigma=0.5$ for the log-normal, such that the average degrees become $7.47\pm 0.30$ and $7.87 \pm 0.05$ ($\pm$ represents the standard deviation over 100 samples of the network with a given degree distribution), respectively. Let $y$ be a hypothetical trait taking values 0 or 1 (e.g.\ being healthy or infected with a disease). We select $15\%$ of the population, starting with the individual with the largest degree and proceeding in decreasing order of degree, to assign the value $y=1$. The remaining individuals in the population are assigned $y=0$. To reduce degree-trait correlations, we go through all vertices and with probability 0.2, we uniformly select a second vertex and swap states (e.g. infected $\rightarrow$ non-infected). This procedure conserves the total number of infected individuals in the population.

\begin{figure}[h]
\centering
\includegraphics[scale=0.8]{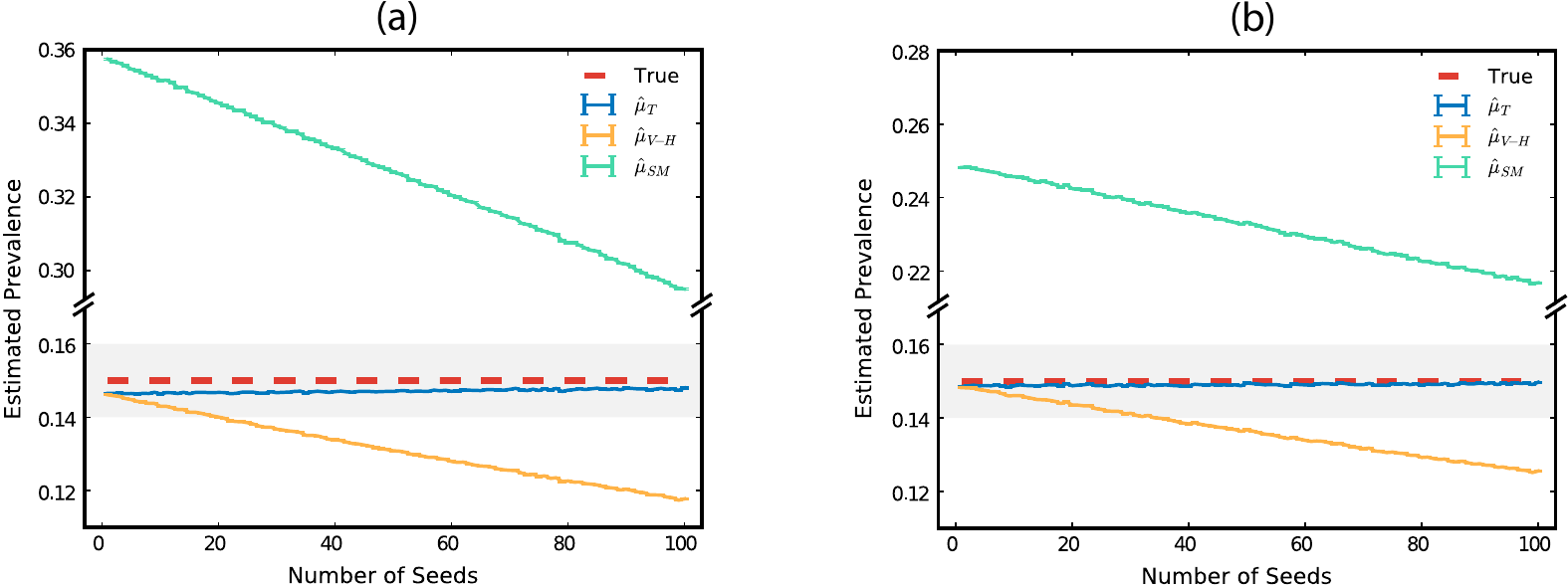}
\caption{Comparative performance of the estimators in a network with a single connected component. Mean of the estimated prevalence and respective standard error (vertical bars, generally smaller than the width of the curves) for varying number of initial seeds (a) Power-law with exponential cutoff and (b) Log-normal degree distributions. We repeat the simulations 100 times for each of the 100 random network samples, therefore, the average and standard error are calculated over 10000 realizations for each number of seeds. Note that the vertical axis is broken in both (a) and (b).}\label{Fig:fig02}
\end{figure}

We start the RDS process with $m$ seeds uniformly chosen within the target population. All seeds start recruitment at the same time. At each time step, an individual invites 3 peers. We assume that all invited peers participate in the experiment. An individual that has already participated in the study may not be invited again at a later time. Recruitment thus stops if the desired sample size $n_S=300$ is achieved or no more recruitments occur. Figure \ref{Fig:fig02} shows the performance of our estimator $\hat\mu_{\text T}$ in comparison to V-H ($\hat\mu_{\text{V-H}} = \frac{\sum_{u\in S}y_u d_u^{-1}}{\sum_{u\in S}d_u^{-1}}$) and the sample mean ($\hat\mu_{\text{SM}} = \frac{1}{n_{S}}\sum_{u\in S} y_u$) for two configurations of networks with power-law with exponential cutoff (Fig.~\ref{Fig:fig02}~a) and log-normal (Fig.~\ref{Fig:fig02}~b) degree distributions. For all estimators, we include the seeds in the sample. Comparatively, our estimator has the best performance irrespective of the number of seeds or degree distribution, slightly underestimating the true prevalence. The V-H estimator increasingly underestimates the true prevalence for increasing number of seeds but performs similarly to our estimator for low number of seeds ($m \lesssim 5$). The sample mean, on the other hand, substantially over-estimates the prevalence as expected but improves performance for increasing number of seeds.

\begin{figure}
\centering
\includegraphics[scale=0.8]{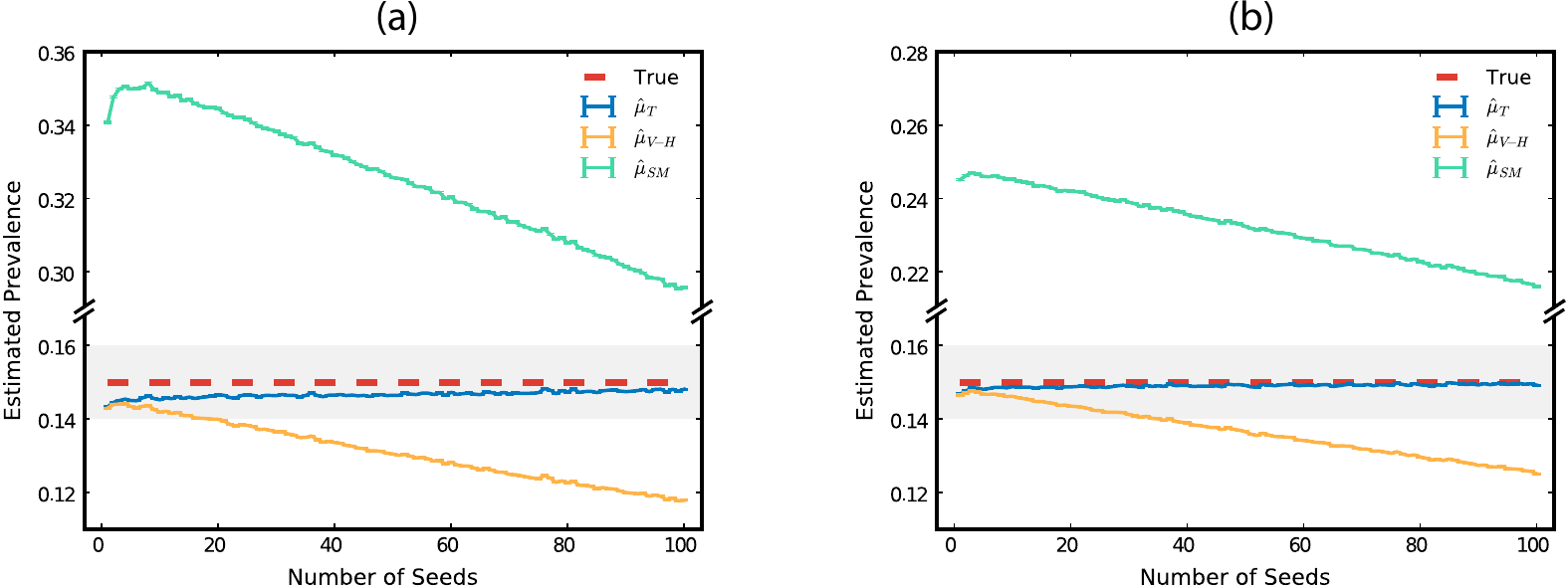}
\caption{Comparative performance of the estimators in a network with two connected components. Mean of the estimated prevalence and respective standard error (vertical bars, generally smaller than the width of the curves) for varying number of initial seeds (a) Power-law with exponential cutoff and (b) Log-normal degree distributions. We repeat the simulations 100 times for each of the 100 random network samples, therefore, the average and standard error are calculated over 10000 realizations for each number of seeds. Note that the vertical axis is broken in both (a) and (b).}\label{Fig:fig03}
\end{figure}

We now make an experiment on a social network with two connected components. We first divide the population into two groups of 5000 vertices each. We then generate stubs for each vertex in the same way as before but only uniformly connect vertices belonging to the same group. The trait y is distributed according to the degree of the vertices, as done for the single component case. The power-law with exponential cutoff now has mean degree $7.43\pm 0.29$ and the log-normal $7.87\pm 0.05$. Although the V-H estimator are not designed for such disconnected networks, in practice one does not know if the social network is connected and thus simply apply the estimator on the collected data. Our estimator however can be safely applied in such settings without restrictions. We nevertheless compare the performance of all three estimators by repeating the previous experiment on this disconnected network. Figure~\ref{Fig:fig03} shows that our estimator generally outperforms V-H and the sample mean. Nevertheless, our estimator slightly underestimates the true prevalence if few seeds are used. The mismatch is larger for the power-law with exponential cutoff network (Fig.~\ref{Fig:fig03}~a) in comparison to the log-normal case (Fig.~\ref{Fig:fig03}~b).

\begin{figure}
\centering
\includegraphics[scale=0.8]{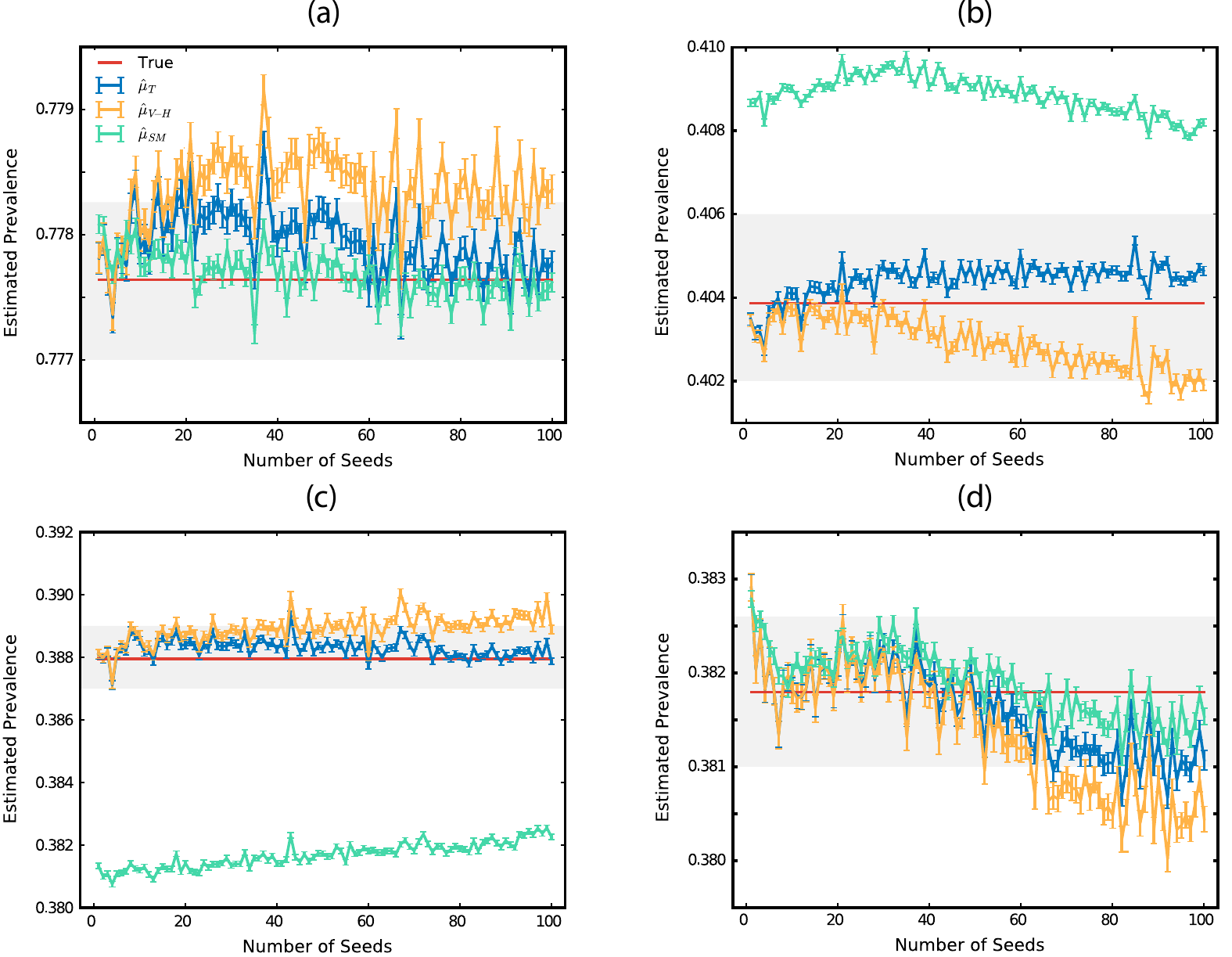}
\caption{Comparative performance of the estimators in a social network with given prevalence of four different traits. Mean of the estimated prevalence and respective standard error (vertical bars) for varying number of initial seeds (a)  Age, (b) Civil Status, (c) County, and (d) Profession. We repeat the simulations 10000 times for different starting conditions, therefore, the average and standard error are calculated over 100000 realizations for each number of seeds.}\label{Fig:fig04}
\end{figure}

We now compare the estimators in a realistic setting, in which the network structure and the prevalence of various individual traits are known. We use an online social network targeting homosexual, bisexual, transgender, and queer persons called Qruiser (www.qx.se); this network was previously analysed in Rybski et al. (2009) and used to evaluate RDS in Lu et al. (2012). The network is connected and contains 16,082 individuals which identify themselves as homosexual males and has 108,334 social ties. The average degree is 13.47. For each individual, 4 dichotomous properties have been extracted from his user profile: age (born before 1980/others), civil status (single/others), county (live in Stockholm/others), and profession (employed/others). Here we also target a sample size of 300 and an individual may invite 3 peers. Figure~\ref{Fig:fig04}~b,c show that our estimator outperforms the other two estimators for detecting the civil status and the county of living, respectively. In these two cases, the correction given by our estimator becomes visible as the number of seeds increases. For the age and profession (Figures~\ref{Fig:fig04}~a,d), on the other hand, our estimator performs similarly to the sample mean but better than the V-H estimator. We see that even in those situations in which V-H performs well, some improvement is obtained by using our estimator.

\section{Discussion}\label{Sec:Discussion}

In this work, we present a novel RDS estimator that utilises a RWWT approximation of the RDS recruitment process. The new estimator is able to account for the multiple seed structure of RDS not considered by the usual simple random walk approximation of RDS. It is also valid for populations with disconnected social networks and does not require information that is traditionally not collected in an RDS sample. To test the performance of our estimator against the V-H estimator and the sample mean, we simulate RDS experiments on theoretical networks with a given prevalence of an hypothetical binary variable $y$. The results show that our estimator generally outperforms the V-H estimator and the sample mean irrespective of the number of seeds. We also performed simulations on a real online social network. In this more complex situation, our estimator overall performs better than the V-H estimator and the sample mean, although the improvement with respect to the number of seeds is not as large as for the generated networks. In our experiments on configuration model networks, the variable $y$ is preferentially distributed in high-degree individuals. In this scenario, both our and the V-H estimators underestimate whereas the sample mean substantially over-estimates the true prevalence. The difference between our estimator and the V-H estimator gets larger for increasing number of seeds, but our estimator performs substantially better. This is expected since the component of the estimator accounting for the assumed simple random sampling of the seeds gets more relevant and thus the performance of V-H decreases significantly with increasing number of seeds. Since it is not uncommon that the seeds correspond to $5-10\%$ of the sample size in empirical studies \citep{malekinejad2008}, or even larger proportions in certain studies, our results show that one may expect substantial biases in the estimates given by the V-H estimator. This bias, generated by the seeds, becomes small or negligible when our estimator is used; additionally, we conclude that the situation with additional seeds is not a major problem for RDS if the corrected estimator is adopted.

In actual RDS practice, seeds are not likely to be selected randomly. Rather, because the seeds are typically chosen among population members known to researchers, the seeds will form a convenience sample, the dependence on which the usual RDS assumption of convergence to equilibrium is meant to handle. However, little or no information exists on the composition of the seeds with respect to sampled properties and network structure in most RDS studies. Nevertheless, uniform sampling is generally a reasonable first approximation. It is often recommended that the seeds are selected such as to reflect the composition of the population \citep{who2013}. The ambition to select a diversified seed sample may result in seeds being selected from the parts of the network that are separate from each other, or that have weak connections. Hence, this ambition may aid in the actual network of coupon distribution not being connected, in which case our estimator is to be preferred.

If the seeds are removed from the sample, the individuals that remain were sampled with probability proportional to degree. Hence, if we estimate population properties without the seeds, we recover the V-H estimator despite that the sample is assumed to come from a RWWT. This implies that the assumption of a connected network is superfluous for the V-H estimator if the seeds are not used for estimation purposes or if the seeds are assumed to be selected with probability proportional to degree. In theory, if the sample is assumed to come from a RWWT, we would need to assume that the social network of the population is a configuration model network in order to use the V-H estimator. However, in the practical estimation process, this becomes a technicality, and we argue that it should not be necessary to make this assumption. The results for other random graph models mentioned in Subsection~\ref{Subsec:StatDist} further supports this argument. 

Following our results, we recommend the use of our estimator: i) if the proportion of seeds in the sample is more than 5\%, either from the initial seeds or from additional seeds that joined along the experiment; ii) if the social network is expected to be disconnected or with weak ties between groups of individuals (e.g. segregated or highly clustered groups inside the target population). Finally, our estimator requires a few more steps for calculation than the well-known V-H estimator. We thus provide a step-by-step guide on how to implement the estimation procedure in the Appendix. Note that no new information is necessary to use our estimator but the number of seeds and degree of the sampled individuals as available in typical RDS studies.

\section*{Acknowledgements}

Jens Malmros is supported by The Swedish Research Council, project no. 621-2012-3868. LECR is a Charg\'e de recherche of the Fonds de la Recherche Scientifique - FNRS. The authors would like to thank Fredrik Liljeros and Xin Lu for use of the Qruiser data and Tom Britton for helpful discussions.

\section*{Appendix: Estimation procedure implementation}

Let $S$ be a respondent-driven sampling (RDS) sample of size $n_S$ from an RDS study with $m$ seeds in which each sampled individual $u$ in $S$ is surveyed for a variable $y_u$ and has degree $d_u$. We proceed as follows to obtain estimates $\hat\mu_{\text T}$ of the mean of $y$.
\begin{enumerate}
\item Calculate an estimate $\hat c$ from Eq.~\eqref{Eq:chat}.
\item Split $S$ into two samples: $S_J$ which consists of the $m$ seeds and $S_{RW}$ which consists of the rest of the sample.
\item Calculate the following estimates:
\begin{enumerate}
\item $\widehat{\expectation(D)}_J$ as the sample mean of the degrees of individuals in $S_J$ from Eq.~\eqref{Eq:EDJ}.
\item $\widehat{\expectation(D)}_{RW}$ as the harmonic mean of the degrees of individuals in $S_{RW}$ from Eq.~\eqref{Eq:EDRW}
\item $\widehat{Var}\left(\widehat{\expectation(D)}_J\right)$ from Eq.~\eqref{Eq:VarJ}. $s_J^2=(1/(m-1))\sum_{u\in S_{J}}(d_u-\bar d_J)^2,$ where $\bar d$ is the mean degree of individuals in $S_J$.
\item $\widehat{Var}\left(\widehat{\expectation(D)}_{RW}\right)$ from Eq.~\eqref{Eq:VarRW}. $s_{d^{-1}}^2=1/(n_S-m-1)\sum_{u\in S_{RW}}(1/d_u-\overbar{d^{-1}})^2,$ where $\overbar{d^{-1}}$ is the mean of $1/d_u$, for sampled $u$ in $S_{RW}$.
\item $\hat w^*$ from Eq.~\eqref{Eq:OptimalWeight} with substituted estimates $\widehat{Var}\left(\widehat{\expectation(D)}_J\right)$ and $\widehat{Var}\left(\widehat{\expectation(D)}_{RW}\right)$.
\item $\widehat{\expectation(D)}$ from Eq.~\eqref{Eq:CompositeEstimator} where we put $w=\hat w^*$.
\end{enumerate}
\item Estimate the draw-wise selection probability $\hat\pi_u$ for every sampled individual $u\in S$ from Eq.~\eqref{Eq:pi_uEstimate}.
\item Estimate the mean of $y$ with the estimator $\hat\mu_{\text T}$ from Eq.~\eqref{Eq:EstRDSEstimator}.
\end{enumerate}

\bibliographystyle{rss}
\bibliography{MasterBibliography}

\end{document}